# A generalisation of the method of regression calibration and comparison with Bayesian and frequentist model averaging methods

Mark P Little[a, b 1], Nobuyuki Hamada[c], Lydia B Zablotska[d]

[a]Radiation Epidemiology Branch, National Cancer Institute, Bethesda, MD 20892-9778 USA

[b]Faculty of Health and Life Sciences, Oxford Brookes University, Headington Campus, Oxford, OX3 0BP, UK

[c]Biology and Environmental Chemistry Division, Sustainable System Research Laboratory, Central Research Institute of Electric Power Industry (CRIEPI), 1646 Abiko, Chiba 270-1194, Japan (ORCID ID 0000-0003-2518-6131)

[d]Department of Epidemiology and Biostatistics, School of Medicine, University of California, San Francisco, 550 16th Street, 2nd floor, San Francisco, CA 94143, USA

[1]**Corresponding author:** Mark P. Little [ORCID ID 0000-0003-0980-7567], Radiation Epidemiology Branch, National Cancer Institute, Room 7E546, 9609 Medical Center Drive, MSC 9778, Rockville, MD 20892-9778 USA, Tel +1 240 276 7375 (office) / +1 301 875 3413 (mobile) / Fax +1 240 276 7840, E-mail address: mark.little@nih.gov

**Word count (main text, excluding Abstract, references):** 4257.

**Word count (abstract):** 445.

**References:** 52

**Supplements:** A and B

**Abstract**

For many cancer sites low-dose risks are not known and must be extrapolated from those observed in groups exposed at much higher levels of dose. Measurement error can substantially alter the dose-response shape and hence the extrapolated risk. Even in studies with direct measurement of low-dose exposures measurement error could be substantial in relation to the size of the dose estimates and thereby distort population risk estimates. Recently, there has been considerable attention paid to methods of dealing with shared errors, which are common in many datasets, and particularly important in occupational and environmental settings.

In this paper we test Bayesian model averaging (BMA) and frequentist model averaging (FMA) methods, the first of these similar to the so-called Bayesian two-dimensional Monte Carlo (2DMC) method, and both fairly recently proposed, against a very newly proposed modification of the regression calibration method, the extended regression calibration (ERC) method, which is particularly suited to studies in which there is a substantial amount of shared error, and in which there may also be curvature in the true dose response. The quasi-2DMC with BMA method performs well when a linear model is assumed, but very poorly when a linear-quadratic model is assumed, with coverage probabilities both for the linear and quadratic dose coefficients that are under 5% when the magnitude of shared Berkson error is large (50%). For the linear model the bias is generally under 10%. However, using a linear-quadratic model it produces substantially biased (by a factor of 10) estimates of both the linear and quadratic coefficients, with the linear coefficient overestimated and the quadratic coefficient underestimated. FMA performs as well as quasi-2DMC with BMA when a linear model is assumed, and generally much better with a linear-quadratic model, although the coverage probability for the quadratic coefficient is uniformly too high. However both linear and quadratic coefficients have pronounced upward bias, particularly when Berkson error is large. By comparison ERC yields coverage probabilities that are too low when shared and unshared Berkson errors are both large (50%), although otherwise it performs well, and coverage is generally better than the quasi-2DMC with BMA or FMA methods, particularly for the linear-quadratic model. The bias of the predicted relative risk at a variety of doses is generally smallest for ERC, and largest for the quasi-2DMC with BMA and FMA methods (apart from unadjusted regression), with standard regression calibration and Monte Carlo maximum likelihood exhibiting bias in predicted relative risk generally somewhat intermediate between ERC and the other two methods. In general ERC performs best in the scenarios presented, and should be the method of choice in situations where there may be substantial shared error, or suspected curvature in the dose response.

**Introduction**

Moderate and high doses of ionising radiation are well established causes of most types of cancer [1,2]. There is emerging evidence, particularly for leukaemia and thyroid cancer, of risk at low dose (<0.1 Gy) radiation[3-6] (roughly 50 times the dose from background radiation in a year). For most other cancer endpoints it is necessary to assess risks via extrapolation from groups exposed at moderate and high levels of dose[7-13]. Such extrapolations, which are dependent on knowing the true dose-response relationship, as inferred from some reference moderate/high-dose data (very often the Japanese atomic bomb survivors), are subject to some uncertainty, not least that induced by systematic and random dosimetric errors that may be present in that moderate/high-dose data[1,14]. Extensive biostatistical research over the last 30 years have done much to develop understanding of this issue[15-30] and in particular the role played by various types of dose measurement error[31]. Among the simplest methods of correction for dose error, regression calibration, which entails substitution of the conditional expectation of the true dose given the observed dose, is straightforward to apply, and is often used to correct for classical error[31]. However, it only takes account of the 1$^{st}$ order dose error terms in the Taylor expansion of the likelihood, and does not take account of correlations between dose errors[32]. It is also prone to bias when dose errors are large, or errors are differential, or the dose response has substantial curvature[33]. Other methods of correction for dose error, in particular Markov Chain maximum likelihood (MCML)[25,26,30,34], and fully Bayesian methods[21-23,29], both of which take full account of the uncertainty in doses (in particular 2$^{nd}$ and higher order dose error terms in the Taylor expansion of the likelihood), can work better in these circumstances.

A variant of the commonly used method of regression calibration[31] has been very recently proposed which is particularly suited to studies with substantial shared error with dose response non-linearity [32]. This so-called extended regression calibration (ERC) method can be used in settings where there is a mixture of Berkson and classical error[32]. In fits to synthetic datasets in which there is substantial upward curvature in the true dose response, and varying (and sometimes substantial) amounts of classical and Berkson error, the ERC method generally outperformed both standard regression calibration, MCML and unadjusted regression, particularly with respect to the coverage probabilities of the quadratic coefficient, and for larger magnitudes of the Berkson error, whether this is shared or unshared[32].

A Bayesian model averaging (BMA) method has been also recently proposed, the so-called 2-dimensional Monte Carlo with Bayesian model averaging (2DMC with BMA) method[28], which has been used in fits to radiation thyroid nodule data[35]. The so-called frequentist model averaging (FMA) model has also been recently proposed, although only fitted to simulated data[36]. In the present paper we shall assess the performance of a variant implementation of the 2DMC with BMA method, which is more closely aligned with standard implementations of BMA[37], and FMA against ERC, making also comparisons with other methods of correction for dose error using simulated data. The simulated data used is exactly as in the previous report[32].

## Methods

### *Synthetic data used for assessing corrections for dose error*

The methods and data exactly parallel those of the previous paper,[32] using publicly available Life Span Study (LSS) leukaemia data[38] to guide construction of a number of artificial datasets. Specifically we used the person year distribution by bone marrow dose groups 0-0.07, 0.08-0.19, 0.20-0.99, 1.00-2.49, ≥2.50 Gy. The central estimates of dose we assumed are close to the person year weighted means of these groups, and as given in Supplement A Table A1, although for the uppermost dose group we assigned a central estimate of 2 Gy. A composite Berkson-classical error model was used in which the true dose $D_{true,i,j}$ and the surrogate dose $D_{surr,i,j}$ to individual $i$ (in dose group $k_i$) in simulation $j$ are given by:

$$D_{true,i,j} = D_{cent,k_i} \exp\left[-0.5(\sigma_{share,Berkson}^2 + \sigma_{unshare,Berkson}^2)\right] \exp\left[\sigma_{share,Berkson}\varepsilon_j + \sigma_{unshare,Berkson}\delta_{i,j}\right] \quad (1)$$

$$D_{surr,i,j} = D_{cent,k_i} \exp\left[-0.5(\sigma_{share,Class}^2 + \sigma_{unshare,Class}^2)\right] \exp\left[\sigma_{share,Class}\mu_j + \sigma_{unshare,Class}\kappa_{i,j}\right] \quad (2)$$

The variables $\varepsilon_j, \delta_{i,j}, \mu_j, \kappa_{i,j}$ are independent identically distributed $N(0,1)$ random variables. It should be noted that the errors $\varepsilon_j, \mu_j$ are common to all individuals and are uniform within each sub-simulation nested within each meta-simulation, and hence give rise to a shared error structure. By comparison the $\delta_{i,j}, \kappa_{i,j}$ are chosen so as to be independent for all other individuals within each sub-simulation and within a meta-simulation, as well as independent of those in other sub-simulations/meta-simulations, and hence give rise to an unshared error structure. The factors $D_{cent,k_i}, D_{cent,k_i}$ are the central estimates of dose, as given in Supplement A Table A1. The factors $\exp\left[-0.5(\sigma_{share,Berkson}^2 + \sigma_{unshare,Berkson}^2)\right]$ and $\exp\left[-0.5(\sigma_{share,Class}^2 + \sigma_{unshare,Class}^2)\right]$ ensure that the distributions given by (1) and (2) have theoretical mean that coincides with the central estimates $D_{cent,k_i}$. The model has the feature that when the Berkson error geometric standard deviations (GSD) are set to 0 ($\sigma_{share,Berkson} = \sigma_{unshare,Berkson} = 0$) the model reduces to one with classical error (a mixture of shared and unshared); likewise when the classical error GSDs are set to 0 ($\sigma_{share,Class} = \sigma_{unshare,Class} = 0$) the model reduces to one with pure Berkson error (a mixture of shared and unshared).

We generated a number of different versions of the dose data, with $\sigma_{share,Berkson}$, $\sigma_{unshare,Berkson}$, $\sigma_{share,Class}$, $\sigma_{unshare,Class}$ taking values of 0.2 (20%) or 0.5 (50%). This individual dose data was then used to simulate the distribution of $N = 250$ cancers for each of $m = 1000$ sub-simulated datasets, using a model in which the assumed probability of being a case for individual $i$ in simulation $j$ ($j = 1,...,m = 1000$) is given by:

$$\exp[\kappa_j][1 + \alpha D_{true,i,j} + \beta D_{true,i,j}^2] \qquad (3)$$

the scaling constant $\kappa_j$ being chosen for each simulation (but not for the Bayesian model fits) to make these sum to 1. As previously, we assumed a linear-quadratic model, with coefficients $\alpha = 0.25 / \text{Gy}, \beta = 2 / \text{Gy}^2$, also a linear model with $\alpha = 3 / \text{Gy}, \beta = 0 / \text{Gy}^2$, both models derived from fitting a stratified linear-quadratic or linear model to the LSS leukaemia data[38].

A total of $n$=500 meta-simulations, consisting of ensembles of dose+cancer simulations were used. Within each meta-simulation a total of $m = 1000$ sub-samples were taken of each type of dose (true dose $= D_{true,i,j}$, surrogate dose $= D_{surr,i,j}$). The true dose ($= D_{true,i,j}$) averaged over the 1000 sub-simulations within each meta-simulation will be used to generate the distribution of cancers by dose group within the meta-simulation, using the summed relative risks $RR_{ij} = 1 + \alpha D_{true,i,j} + \beta D_{true,i,j}^2$. Within each of the $n$=500 meta-simulations, this cancer distribution will be constant over the $m = 1000$ sub-simulations that comprise it, but this distribution will of course change slightly between each of the $n = 500$ meta-simulations. The $n = 500$ meta-simulated dose+cancer ensembles were used to fit models and evaluate fitted model means and coverage probability. Having derived synthetic individual level data, for the purposes of model fitting, for all models except MCML and 2DMC with BMA, the data were then collapsed (summing cases, averaging doses) into the 5 dose groups used previously[32].

Poisson linear relative risk generalised linear models[39] were fitted to this grouped data, with rates given by expression (3), using as offsets the previously-specified number per group[32]. Models were fitted using six separate methods (unadjusted, regression calibration, ERC, MCML, 2DMC and BMA, FMA). For ERC and the other methods previously used the methods of deriving doses and model fitting were as in our earlier paper[32]. It should be noted that for all except the unadjusted regressions the true (rather than surrogate) dose is used, so that classical error does not figure. Only for unadjusted regression is it the surrogate dose that is used.

We used a BMA method somewhat analogous to the 2DMC with BMA method of Kwon *et al*[28], using the full set of mean true doses per group as previously generated for MCML, the mean doses per group for each simulation being given by group means of the samples generated by expression (3), averaged over the $m = 1000$ dose samples. The model was fitted using Bayesian Markov Chain Monte Carlo (MCMC) methods. Associated with the dose vector is a vector of probabilities $p_j, j = 1,...,1000$ which is generated using variables $\lambda_j, j = 1,...,999$, so that:

$$p_j = \frac{\exp[\lambda_j]}{1 + \sum_{k=1}^{999} \exp[\lambda_k]}, j = 1,...,999, \quad p_{1000} = \frac{1}{1 + \sum_{k=1}^{999} \exp[\lambda_k]} \qquad (4)$$

This is therefore quite close to the method proposed by Hoeting *et al*[37], and somewhat distinct from the formulation of 2DMC with BMA proposed by Kwon *et al*[28], as we discuss at greater length below. For this reason we shall describe our own method as quasi-2DMC with BMA. The standard formulation of BMA, as given by Hoeting *et al*[37], and which we employ, is based on the posterior probability:

$$p(\alpha,\beta,\kappa\,|\,Y,X,(D_{true,k})_{k=1}^{m},(p_{j})_{k=1}^{m})$$

$$=\sum_{k=1}^{m}p(\alpha,\beta,\kappa\,|\,\gamma=k,Y,X,D_{true,\gamma})p(\gamma=k\,|\,Y,X,(D_{true,k})_{k=1}^{m},(p_{j})_{k=1}^{m}) \qquad (5)$$

$$=\sum_{k=1}^{m}p(\alpha,\beta,\kappa\,|\,Y,X,D_{true,k})p_{k}$$

where $(p_k)_{k=1}^{m}$ are given by Eq. (4). We fitted via successive application of Metropolis-Hastings samplers to (a) sample conditionally the model dose response parameters $\alpha,\beta,\kappa$ conditionally on $(p_k)_{k=1}^{m}$ and (b) sample conditionally the dose vector probability parameters $(\lambda_j)_{k=1}^{m-1}$ which determine $(p_k)_{k=1}^{m}$ conditionally on $\alpha,\beta,\kappa$. Kwon et al[28] did this slightly differently, sampling (a) the dose response parameters $\alpha,\beta,\kappa$ conditional on $k$, (b) sampling of $k$ (via a multinomial distribution) conditional on $\alpha,\beta,\kappa$ and $(p_k)_{k=1}^{m}$ and (c) sampling of the parameters $(p_k)_{k=1}^{m}$ (via a Dirichlet distribution) conditional on $\alpha,\beta,\kappa$ and $k$. Kwon et al[28] resorted to use of an approximate Monte Carlo sampler, the so-called stochastic approximation Monte Carlo (SAMC) method of Liang et al [40]. Unfortunately Kwon et al do not provide enough information to infer the precise form of SAMC that was used by them[28], and for that reason we have adopted this alternative.

All main model parameters ($\alpha,\beta,\kappa,\lambda_k$) had normal priors, with mean 0 and standard deviation (SD) 1000. The Metropolis-Hastings algorithm was used to generate samples from the posterior distribution. Random normal proposal distributions were assumed for all variables, with SD of 0.2 for $\kappa$ and 1 for $\alpha,\beta$, and SD 2 for all $\lambda_k$. The $\lambda_k$ were proposed in blocks of 10. Two separate chains were used, in order to compute the Brooks-Gelman-Rubin (BGR) convergence statistic [41,42]. The first 1000 simulations were discarded, and a further 1000 simulations taken for sampling. The proposal SDs and number of burn-in sample were chosen to give mean BGR statistics (over the 500 simulated datasets) that were in all cases less than 1.03 and acceptance probabilities of about 30% for the main model parameters ($\alpha, \beta, \kappa$), suggesting good mixture and likelihood of chain convergence. For the ERC model confidence intervals were (as previously) derived using the profile likelihood [39] and for the quasi-2DMC with BMA model Bayesian uncertainty intervals were derived.

The FMA model of Kwon et al[36] was also fitted. For each of $j=1,...,m=1000$ dose vectors fits model (3) using Poisson regression (via maximum likelihood) and the Akaike Information Criterion (AIC)[43], $AIC_j$, computed, as well as the central estimate and profile likelihood 95% confidence intervals for each coefficient, $\alpha_{j,MLE}(\alpha_{j,0.025}, \alpha_{j,0.975})$ and $\beta_{j,MLE}(\beta_{j,0.025}, \beta_{j,0.975})$; from these were derived the estimated standard deviation, via

$$SD(\alpha)_j = \min(\alpha_{j,MLE} - \alpha_{j,0.025}, \alpha_{j,0.975} - \alpha_{j,MLE})/1.96$$

and

$$SD(\beta)_j = \min(\beta_{j,MLE} - \beta_{j,0.025}, \beta_{j,0.975} - \beta_{j,MLE})/1.96$$

, in other words the minimum of the distance from each CI to the mean, divided by the asymptotic 97.5% point (1.96) of the normal distribution, to recover the SD. For each fit $k=100$ simulations were taken from the respective normal distributions $N(\alpha_{j,MLE}, SD(\alpha)_j^2)$ and $N(\beta_{j,MLE}, SD(\beta)_j^2)$, and each such sample given an AIC-derived weight via:

$$\exp[-AIC_j/2]/\sum_{k=1}^{1000}\exp[-AIC_k/2] \qquad (6)$$

[It should be noted that Eq. (6) differs from the formula mistakenly given by Kwon et al[36] in their paper, $\exp[AIC_j/2]/\sum_{k=1}^{1000}\exp[AIC_k/2]$.] The central estimate for each coefficient was taken as the AIC-derived-weighted sum of these samples, and 95% CI estimated from the 2.5% and 97.5% centiles of the AIC-derived-weighted samples. A variety of $k$ in the range 100-1000 were used, yielding very similar results. We also tried using asymmetric confidence intervals, employing $SD(\alpha)_{j,lower} = (\alpha_{j,MLE} - \alpha_{j,0.025})/1.96$ and $SD(\alpha)_{j,upper} = (\alpha_{j,0.975} - \alpha_{j,MLE})/1.96$ to separately generate the samples above and below the central estimate $\alpha_{j,MLE}$, and likewise for the $\beta$ coefficient. However, this generally yielded badly biased estimates of both coefficients, because of occasional samples in which one or other CI was very large.

The Fortran 95-2003 program used to generate these datasets and perform Poisson and Bayesian MCMC model fitting, and the relevant steering files employed to control this program are given in online Supplement B. Using the mean coefficients for each model and error scenario over the 500 simulated datasets, $(\alpha_{mean}, \beta_{mean})$, the percentage mean bias in predicted excess relative risk (ERR) is calculated, via:

$$100\left[(\alpha_{mean}D_{pred} + \beta_{mean}D_{pred}^2)/(\alpha D_{pred} + \beta D_{pred}^2) - 1\right] \tag{7}$$

This was evaluated for two values of predicted dose, $D_{pred} = 0.1$ Gy and $D_{pred} = 1$ Gy.

*Data availability statement*

The datasets generated and analysed in the current study are available by running the Fortran 95/2003 program **fitter_shared_error_simulation_reg_cal_Bayes_FMA.for**, given in the online web repository, with any of the 12 steering input files given there. All are described in Supplement B. The datasets are temporarily stored in computer memory, and the program uses them for fitting the Poisson models described in the Methods section.

**Results**

As shown in Table 1, using the linear-quadratic model the coverage probabilities of the ERC method for the linear coefficient $\alpha$ are near the desired 95% level, irrespective of the magnitudes of assumed Berkson error, whether shared or unshared. However, the ERC method yields coverage probabilities that are somewhat too low when shared and unshared Berkson errors are both large (with logarithmic SD=50%), although otherwise it performs well (Table 1). It should be noted that classical error will have no effect on any of these models, as its only effect is on the unadjusted regression model (via sampling of the surrogate dose), and so the effect is not shown. By contrast the coverage probabilities of both the linear coefficient $\alpha$ and the quadratic coefficient $\beta$ for the quasi-2DMC with BMA method are generally much too low, and when shared Berkson error is large (50%) the coverage probabilities do not exceed 5% (Table 1). The coverage for the FMA method is generally better, and for the coefficient $\alpha$ does not depart too markedly from the desired 95%; however, the coverage of the coefficient $\beta$ tends to be too high, for any non-zero level of Berkson error, whether shared or unshared (Table 1).

Table 2 shows that for the linear model the coverage percentage is generally too high for ERC, MCML and FMA, and slightly too low for quasi-2DMC with BMA, but more or less correct for regression calibration.

Table 3 shows the coefficient mean values, averaged over all 500 simulations, assuming a linear-quadratic model. A notable feature is that for larger values of Berkson error, the linear coefficient $\alpha$ for quasi-2DMC with BMA is substantially overestimated, and the quadratic dose coefficient $\beta$ substantially underestimated, both by factors of about 10. For ERC the estimates of the quadratic coefficient $\beta$ are upwardly biased, but not by such large amounts. For FMA both coefficients have pronounced upward bias, particularly for large shared Berkson error (50%) (Table 3).

Table 4 shows that the bias in ERR for ERC evaluated either at 0.1 Gy or 1 Gy does not exceed 30% in absolute value. Regression calibration performs somewhat worse, with bias ~60% when shared and unshared Berkson error are large (50%) for predictions at 1 Gy, although otherwise with bias under 30%. For all but the smallest shared and unshared Berkson errors (both 0% or 20%) MCML has bias ~35-70%, with bias particularly severe at 1 Gy. Quasi-2DMC with BMA performs somewhat worse, with bias in excess of ~30% and sometimes in excess of 100% when Berkson errors are large, and bias most pronounced at low dose. Unadjusted regression yields almost as severe bias, which exceeds 50% in many cases for predictions at 1 Gy, and when shared or unshared classical errors are large (50%) often exceeds 100% (Table 4). Bias for FMA is generally moderate to severe, and particularly bad (>100%) when shared Berkson error is large (50%) (Table 4).

Table 5 shows the coefficient mean values, averaged over all 500 simulations, for the linear model, and percentage bias. For most methods bias is modest, generally under ~10%. The only significant exception is FMA where the bias approaches 30% when shared Berkson errors are large (50%) (Table 5).

It should be noted that the behaviour of all regression methods in both scenarios (linear, linear-quadratic) is very similar when Berkson errors are 50%, irrespective of whether unshared Berkson errors are 0, 20% or 50% (Tables 1-5).

## Discussion

We have demonstrated that the quasi-2DMC with BMA method performs well when a linear model is assumed and fitted, albeit with coverage for the linear coefficient that is slightly too low (~90%) (Table 2). However, it performs very poorly when a linear-quadratic model is assumed and fitted, with coverage probabilities both for the linear and quadratic dose coefficients that are under 5% when the magnitude of shared Berkson errors is large (50%) (Table 1). For the linear model the bias is generally modest, under 10% (Table 5). However, if a linear-quadratic model is assumed there is substantial bias (by a factor of 10) in estimates of both the linear and quadratic coefficients, with the linear coefficient overestimated and the quadratic coefficient underestimated (Table 3). FMA performs as well as the other method when a linear model is assumed (Table 2), and generally much better assuming a linear-quadratic model, although the coverage probability for the quadratic coefficient is uniformly too high (Table 1). However both linear and quadratic coefficients have pronounced upward bias, particularly when Berkson error is large (50%) (Table 3). By comparison the ERC method yields coverage probabilities that are too high when a linear model is fitted (Table 2), and too low when a linear-quadratic model is fitted with shared and unshared Berkson errors both large (50%) (Table 1), although otherwise it performs well, and coverage is generally better than for quasi-2DMC with BMA or FMA, particularly for the linear-quadratic model. As shown previously it generally outperforms all other methods (regression calibration, MCML, unadjusted regression) when a linear-quadratic model is assumed[32]. The upward bias in estimates of the $\alpha$ coefficient and the downward bias in the estimates of the $\beta$ coefficient, at least for larger magnitudes of error (Table 3) largely explains the poor coverage of quasi-2DMC with BMA in these cases (Table 1). The bias of the predicted ERR using the linear-quadratic model at a variety of doses is generally smallest for ERC, and largest (apart from unadjusted regression) for quasi-2DMC with BMA and for FMA, with standard regression calibration and MCML exhibiting bias in predicted ERR generally somewhat intermediate between the other two methods (Table 4). The fact that bias in ERR tends to be larger for a dose of 1 Gy (Table 4, Table 5) relates to the fact that at higher assumed dose the contribution of the quadratic coefficient is relatively more important. However, this is not always the case, and for example for the quasi-2DMC with BMA method the inflated linear coefficient and much reduced quadratic coefficient (Table 3) generally result in bias being more severe at the lower dose (Table 4).

As noted above, the form of the quasi-2DMC with BMA model that we fit differs slightly from that employed by Kwon *et al*[28]. Our method and that of Kwon *et al*[28] should be approximately equivalent, although the latter is considerably more computationally challenging, and may be the reason why Kwon *et al*[28] resorted to use of the SAMC method[40], in order to get their method to work. As noted in the Methods, unfortunately Kwon *et al* do not provide enough information to infer the precise form of SAMC that was used by them[28], and it was for that reason we adopted this alternative, which is in any case possibly more computationally efficient. It is possible that the SAMC implementation used by Kwon *et al*[28] may behave differently from the more standard implementation of BMA given here. Kwon *et al*[28] report results of a simulation study that tested the 2DMC with BMA method against what they term "conventional regression", which may have been regression calibration. They did not assess performance against MCML, and in all cases only a linear model was tested[28] unlike the simulations given here and in a previous publication[32], which used a linear-quadratic model, and in the present paper also a linear model. Kwon *et al*[28] report generally better performance of 2DMC with BMA against the regression calibration alternative. Their findings using the linear model are consistent with ours (Table 2, Table 5). Kwon *et al*[36] tested FMA against the 2DMC method and against the so-called corrected information matrix (CIM) method[44] and observed similar performance, in particular adequate coverage, of all three methods, although narrower CI were produced by FMA compared with CIM under a number of scenarios. However, in all cases only a linear model was tested[36]. Set against this, Stram *et al* reported results of a simulation study[45] which suggested that the 2DMC with BMA method will produce substantially upwardly biased estimates of risk, also that the coverage may be poor, somewhat confirming our own findings, at least using the linear-quadratic model; although we do not always find upward bias, and generally little bias when a linear model is assumed, the coverage of both regression coefficients for quasi-2DMC with BMA is poor for larger values of shared Berkson error when a linear-quadratic model is assumed (Table 1, Table 3, Table 4). One possible reason for the bias that may occur in quasi-2DMC with BMA is that by chance a dose vector is chosen which results in good fit by the model (linear or linear-quadratic) but which nevertheless is substantially biased, resulting in substantial bias in the linear and/or quadratic coefficients. While in most circumstances (as in the present simulations) there is no information regarding the dose vector, it might occur that one has an informative prior for the $p_j$, which would be expected to to reduce the likelihood of such bias.

Dose error in radiation studies is unavoidable, even in experimental settings. It is particularly common in epidemiological studies, in particular those of occupationally exposed groups, where shared errors, resulting from group assignments of dose, dosimetry standardizations or possible variability resulting from application of, for example, biokinetic or environmental or biokinetic models result in certain shared unknown (and variable) parameters between individuals or groups. There have been extensive assessments of uncertainties in dose in these settings[46-48]. Methods for taking account of such uncertainties cannot always correct for them, but they at least enable error adjustment (e.g. to CI) to be made[31]. As previously discussed[32] the defects in the standard type of regression calibration are well known, in particular that the method can break down when dose error is substantial[31], as it is in many of our scenarios. It also fails to take account of shared errors. Perhaps because of this a number of methods have been recently developed that take shared error into account, in particular the 2DMC and BMA method[28] and the CIM method[44]. The CIM method only applies to situations where there is pure Berkson error. Both 2DMC with BMA and CIM have been applied in number of settings, the former to analysis of thyroid nodules in nuclear weapons test exposed individuals[35], and the latter to assessment of lung cancer risk in Russian Mayak nuclear workers[49] and cataract risk in the US Radiologic Technologists[50]. In principle the simulation extrapolation (SIMEX) method[51] can be applied in situations where there is shared (possibly combined with unshared) classical error, where the magnitudes of shared and unshared error are known. However, this was not part of the original formulation of SIMEX[51]. Possibly because of the restrictions on error structure and its extreme computational demands SIMEX has only rarely been used in radiation settings[27,52].

## Conclusions

Using methods and data that exactly parallel those of the previous paper,[32] differing only in that linear as well as linear-quadratic models were assumed, we have demonstrated that the quasi-2DMC with BMA method performs well when a linear model is assumed (Table 2), but very poorly when a linear-quadratic dose response is assumed, with coverage probabilities both for the linear and quadratic dose coefficients that are under 5% when the magnitude of shared Berkson error is moderate to large (Table 1). The bias with a linear model for this method is generally modest (under 10%) (Table 4), but for the linear-quadratic model bias is substantial (by a factor of 10) both for the linear and quadratic coefficients, with the linear coefficient overestimated and the quadratic coefficient underestimated (Table 3). FMA performs generally better, although when assuming a linear-quadratic model the coverage probability for the quadratic coefficient is uniformly too high (Table 1), as it is also for the linear coefficient assuming a linear model (Table 2). However both linear and quadratic coefficients using FMA have pronounced upward bias, particularly when Berkson error is large (50%) (Table 3, Table 4,

Table 5). By comparison the recently developed ERC method[32] yields coverage probabilities that are too high when a linear model is assumed, and too low when a linear-quadratic model is assumed and shared and unshared Berkson errors are both large (50%), although otherwise it performs well, and coverage is generally better than for quasi-2DMC with BMA or FMA, particularly for the linear-quadratic model. The bias of the predicted ERR at a variety of doses is generally smallest for ERC, and largest for quasi-2DMC with BMA and FMA, with standard regression calibration and MCML exhibiting bias in predicted ERR generally somewhat intermediate between the other two methods (Table 4, Table 5). In general ERC performs best in the scenarios presented, and should be the method of choice in situations where there may be substantial shared error, or suspected curvature in the dose response.

## Acknowledgements

The authors are grateful for the detailed and helpful comments of Dr Jay Lubin and the three referees. The Intramural Research Program of the National Institutes of Health, the National Cancer Institute, Division of Cancer Epidemiology and Genetics supported the work of MPL. The work of LBZ was supported by National Cancer Institute and National Institutes of Health (Grant No. R01CA197422). The funders had no role in considering the study design or in the collection, analysis, interpretation of data, writing of the report, or decision to submit the article for publication.

## Contributorship

M.P.L. formulated the analysis, wrote and ran the analysis code and wrote the first draft of the paper. L.B.Z. and N.H. contributed to extensive rewrites of the subsequent drafts of the paper. All authors reviewed the manuscript and approved its submission.

**Table 1. Coverage probability of profile likelihood (for extended regression calibration) or Bayesian posterior confidence intervals (for quasi-2DMC with BMA) or frequentist model averaging (FMA) for fits of linear-quadratic model.** Coverage probability evaluated using $n=500$ dose+cancer ensembles, all with classical error (shared and unshared) of 20%. The central three columns are mostly taken from the paper of Little *et al*[32].

| Magnitude of unshared Berkson error | Magnitude of shared Berkson error | Sample Pearson correlation coefficient between individual true doses | Extended regression calibration adjusted Coverage % | | Quasi-2DMC with BMA Coverage % | | Frequentist model averaging (FMA) Coverage % | |
|---|---|---|---|---|---|---|---|---|
| | | | $\alpha$ | $\beta$ | $\alpha$ | $\beta$ | $\alpha$ | $\beta$ |
| 0% | 0% | NA | 95.2 | 94.8 | 93.6 | 95.0 | 92.6 | 94.8 |
| 20% | 20% | 0.50 | 94.8 | 98.4 | 92.4 | 94.2 | 92.6 | 99.2 |
| 20% | 50% | 0.84 | 95.4 | 94.8 | 4.6 | 1.6 | 91.8 | 100.0 |
| 0% | 50% | 1.00 | 95.8 | 95.4 | 4.6 | 1.6 | 91.8 | 100.0 |
| 50% | 20% | 0.15 | 94.4 | 94.8 | 88.0 | 96.8 | 92.2 | 99.2 |
| 50% | 50% | 0.45 | 95.0 | 80.4 | 0.8 | 3.8 | 94.2 | 100.0 |

**Table 2. Coverage probability of profile likelihood (for extended regression calibration) or Bayesian posterior confidence intervals (for quasi-2DMC with BMA) or frequentist model averaging (FMA) for fits of linear model.** Coverage probability evaluated using $n$=500 dose+cancer ensembles, all with classical error (shared and unshared) of 20%.

| | | Coverage % | | | | | |
|---|---|---|---|---|---|---|---|
| Magnitude of unshared Berkson error | Magnitude of shared Berkson error | Unadjusted | Regression calibration | Extended regression calibration adjusted | Monte Carlo maximum likelihood | Quasi-2DMC with BMA | Frequentist model averaging (FMA) |
| 0% | 0% | 82.4 | 94.8 | 94.8 | 94.8 | 93.8 | 90.4 |
| 20% | 20% | 82.4 | 94.4 | 99.6 | 99.2 | 94.4 | 98.8 |
| 20% | 50% | 81.8 | 95.2 | 99.8 | 100.0 | 91.0 | 100.0 |
| 0% | 50% | 81.8 | 95.2 | 99.8 | 100.0 | 90.8 | 100.0 |
| 50% | 20% | 82.2 | 94.4 | 99.6 | 99.4 | 94.6 | 98.8 |
| 50% | 50% | 81.8 | 95.2 | 99.8 | 100.0 | 91.2 | 100.0 |

**Table 3. Mean over *n*=500 dose+cancer ensembles of regression coefficients in fits of linear-quadratic model, all with classical error (shared and unshared) of 20%.** The central two columns are mostly taken from the paper of Little *et al*[32].

| Magnitude of unshared Berkson error | Magnitude of shared Berkson error | Extended regression calibration | | Quasi-2DMC with BMA | | Frequentist model averaging (FMA) | |
|---|---|---|---|---|---|---|---|
| | | ERR/Gy $\alpha$ | ERR/Gy$^2$ $\beta$ | ERR/Gy $\alpha$ | ERR/Gy$^2$ $\beta$ | ERR/Gy $\alpha$ | ERR/Gy$^2$ $\beta$ |
| 0% | 0% | 0.196 | 2.061 | 0.362 | 2.072 | 0.196 | 2.061 |
| 20% | 20% | 0.125 | 2.132 | 0.891 | 1.672 | 0.205 | 2.516 |
| 20% | 50% | 0.109 | 2.393 | 2.902 | 0.180 | 0.294 | 5.730 |
| 0% | 50% | 0.114 | 2.350 | 2.786 | 0.181 | 0.277 | 5.535 |
| 50% | 20% | 0.121 | 2.354 | 1.108 | 1.971 | 0.247 | 3.060 |
| 50% | 50% | 0.038 | 2.795 | 3.527 | 0.210 | 0.282 | 7.106 |
| **True value** | | **0.25** | **2.0** | **0.25** | **2.0** | **0.25** | **2.0** |

Notes: ERR, excess relative risk.

**Table 4. Percentage bias in excess relative risk (ERR) using linear-quadratic model predicted at 0.1 Gy or 1 Gy with various sorts of dose error correction, using mean regression coefficients over $n$=500 dose+cancer ensembles (as in Table 2)**

| Magnitude of error distribution | | | | Percentage bias in excess relative risk estimates by type of dose error correction | | | | | | | | | | | |
|---|---|---|---|---|---|---|---|---|---|---|---|---|---|---|---|
| Unshared Berkson error | Shared Berkson error | Unshared classical error | Shared classical error | Unadjusted | | Regression calibration | | Extended regression calibration | | Monte Carlo maximum likelihood | | Quasi-2DMC with BMA | | Frequentist model averaging (FMA) | |
| | | | | ERR at 0.1 Gy | ERR at 1 Gy | ERR at 0.1 Gy | ERR at 1 Gy | ERR at 0.1 Gy | ERR at 1 Gy | ERR at 0.1 Gy | ERR at 1 Gy | ERR at 0.1 Gy | ERR at 1 Gy | ERR at 0.1 Gy | ERR at 1 Gy |
| 0% | 0% | 20% | 20% | -0.26 | 11.07 | -10.58 | 0.34 | -10.58 | 0.34 | -10.58 | 0.34 | 26.49 | 8.17 | -80.13 | 0.33 |
| 0% | 0% | 20% | 50% | 56.68 | 98.07 | -10.58 | 0.34 | -10.58 | 0.34 | -10.58 | 0.34 | 26.49 | 8.17 | -80.13 | 0.33 |
| 0% | 0% | 50% | 20% | 6.99 | 11.79 | -10.58 | 0.34 | -10.58 | 0.34 | -10.58 | 0.34 | 26.49 | 8.17 | -80.13 | 0.33 |
| 0% | 0% | 50% | 50% | 64.85 | 98.42 | -10.58 | 0.34 | -10.58 | 0.34 | -10.58 | 0.34 | 26.49 | 8.17 | -80.13 | 0.33 |
| 20% | 20% | 20% | 20% | 3.74 | 19.52 | -7.04 | 7.90 | -24.73 | 0.34 | 13.14 | 10.91 | 135.05 | 13.88 | -77.43 | 20.97 |
| 20% | 20% | 20% | 50% | 64.26 | 113.79 | -7.04 | 7.90 | -24.73 | 0.34 | 13.14 | 10.91 | 135.05 | 13.88 | -77.43 | 20.97 |
| 20% | 20% | 50% | 20% | 11.14 | 20.28 | -7.04 | 7.90 | -24.73 | 0.34 | 13.14 | 10.91 | 135.05 | 13.88 | -77.43 | 20.97 |
| 20% | 20% | 50% | 50% | 72.64 | 114.21 | -7.04 | 7.90 | -24.73 | 0.34 | 13.14 | 10.91 | 135.05 | 13.88 | -77.43 | 20.97 |
| 20% | 50% | 20% | 20% | 24.61 | 44.25 | 10.67 | 30.42 | -22.65 | 11.21 | 49.11 | 50.12 | 548.93 | 36.99 | -57.18 | 167.75 |
| 20% | 50% | 20% | 50% | 99.33 | 156.92 | 10.67 | 30.42 | -22.65 | 11.21 | 49.11 | 50.12 | 548.93 | 36.99 | -57.18 | 167.75 |
| 20% | 50% | 50% | 20% | 33.08 | 45.12 | 10.67 | 30.42 | -22.65 | 11.21 | 49.11 | 50.12 | 548.93 | 36.99 | -57.18 | 167.75 |
| 20% | 50% | 50% | 50% | 108.90 | 157.37 | 10.67 | 30.42 | -22.65 | 11.21 | 49.11 | 50.12 | 548.93 | 36.99 | -57.18 | 167.75 |
| 0% | 50% | 20% | 20% | 19.81 | 39.10 | 6.25 | 25.80 | -22.52 | 9.49 | 43.65 | 45.57 | 523.08 | 31.85 | -59.00 | 158.29 |
| 0% | 50% | 20% | 50% | 91.80 | 147.91 | 6.25 | 25.80 | -22.52 | 9.49 | 43.65 | 45.57 | 523.08 | 31.85 | -59.00 | 158.29 |
| 50% | 20% | 20% | 20% | 23.79 | 44.18 | 9.93 | 30.36 | -20.88 | 9.97 | 34.39 | 34.02 | 189.99 | 36.82 | -72.69 | 46.98 |
| 50% | 20% | 20% | 50% | 98.02 | 157.23 | 9.93 | 30.36 | -20.88 | 9.97 | 34.39 | 34.02 | 189.99 | 36.82 | -72.69 | 46.98 |
| 50% | 20% | 50% | 20% | 32.41 | 45.02 | 9.93 | 30.36 | -20.88 | 9.97 | 34.39 | 34.02 | 189.99 | 36.82 | -72.69 | 46.98 |
| 50% | 20% | 50% | 50% | 107.78 | 157.59 | 9.93 | 30.36 | -20.88 | 9.97 | 34.39 | 34.02 | 189.99 | 36.82 | -72.69 | 46.98 |
| 50% | 50% | 20% | 20% | 36.22 | 75.38 | 20.91 | 58.16 | -29.45 | 25.90 | 56.13 | 67.27 | 688.36 | 66.08 | -50.97 | 228.37 |
| 50% | 50% | 20% | 50% | 123.65 | 214.48 | 20.91 | 58.16 | -29.45 | 25.90 | 56.13 | 67.27 | 688.36 | 66.08 | -50.97 | 228.37 |
| 50% | 50% | 50% | 20% | 45.28 | 76.32 | 20.91 | 58.16 | -29.45 | 25.90 | 56.13 | 67.27 | 688.36 | 66.08 | -50.97 | 228.37 |
| 50% | 50% | 50% | 50% | 133.97 | 214.90 | 20.91 | 58.16 | -29.45 | 25.90 | 56.13 | 67.27 | 688.36 | 66.08 | -50.97 | 228.37 |

**Table 5. Mean and percentage bias over *n*=500 dose+cancer ensembles of regression coefficients in fits of linear model, all with classical error (shared and unshared) of 20%.**

| Magnitude of unshared Berkson error | Magnitude of shared Berkson error | Unadjusted | | Regression calibration | | Extended regression calibration adjusted | | Monte Carlo maximum likelihood | | Quasi-2DMC with BMA | | Frequentist model averaging (FMA) | |
|---|---|---|---|---|---|---|---|---|---|---|---|---|---|
| | | ERR/Gy | % bias | ERR/Gy | % bias | ERR/Gy | % bias | ERR/Gy | % bias | ERR/Gy | % bias | ERR/Gy | % bias |
| 0% | 0% | 3.124 | 4.14 | 3.019 | 0.63 | 3.019 | 0.63 | 3.019 | 0.63 | 3.150 | 5.00 | 3.019 | 0.63 |
| 20% | 20% | 3.123 | 4.11 | 3.018 | 0.61 | 3.066 | 2.21 | 3.033 | 1.09 | 3.082 | 2.74 | 3.142 | 4.72 |
| 20% | 50% | 3.125 | 4.16 | 3.020 | 0.67 | 3.017 | 0.57 | 3.345 | 11.50 | 2.755 | -8.18 | 3.877 | 29.22 |
| 0% | 50% | 3.125 | 4.15 | 3.020 | 0.66 | 3.017 | 0.57 | 3.349 | 11.63 | 2.752 | -8.27 | 3.876 | 29.20 |
| 50% | 20% | 3.124 | 4.12 | 3.019 | 0.62 | 3.063 | 2.09 | 3.033 | 1.11 | 3.081 | 2.69 | 3.143 | 4.77 |
| 50% | 50% | 3.125 | 4.17 | 3.021 | 0.68 | 3.016 | 0.53 | 3.349 | 11.64 | 2.749 | -8.35 | 3.878 | 29.27 |
| **True value** | | **3.0** | | **3.0** | | **3.0** | | **3.0** | | **3.0** | | **3.0** | |

Notes: ERR, excess relative risk.

44. Zhang, Z. *et al.* Correction of confidence intervals in excess relative risk models using Monte Carlo dosimetry systems with shared errors. *PloS one* **12**, e0174641 (2017). https://doi.org/10.1371/journal.pone.0174641
45. Simon, S. L., Hoffman, F. O. & Hofer, E. Letter to the Editor Concerning Stram et al.: "Lung Cancer in the Mayak Workers Cohort: Risk Estimation and Uncertainty Analysis." Radiat Res 2021; 195:334-46. *Radiation Research* **196**, 449-451 (2021). https://doi.org/10.1667/rade-21-00106.1
46. National Council on Radiation Protection and Measurements (NCRP). *NCRP Report No. 158. Uncertainties in the measurement and dosimetry of external radiation.*, i-xx+1-546 (National Council on Radiation Protection and Measurements (NCRP). 2007).
47. National Council on Radiation Protection and Measurements (NCRP). i-xxii+1-841 (National Council on Radiation Protection and Measurements (NCRP), 7910 Woodmont Avenue, Suite 400 / Bethesda, MD 20814-3095, USA, 2009).
48. National Council on Radiation Protection and Measurements (NCRP). *NCRP Report No. 171. Uncertainties in the estimation of radiation risks and probability of disease causation.*, i-xv+1-418 (National Council on Radiation Protection and Measurements (NCRP). 2012).
49. Stram, D. O. *et al.* Lung Cancer in the Mayak Workers Cohort: Risk Estimation and Uncertainty Analysis. *Radiat Res* **195**, 334-346 (2021). https://doi.org/10.1667/RADE-20-00094.1
50. Little, M. P., Patel, A., Hamada, N. & Albert, P. Analysis of cataract in relationship to occupational radiation dose accounting for dosimetric uncertainties in a cohort of U.S. radiologic technologists. *Radiat. Res.* **194**, 153-161 (2020). https://doi.org/10.1667/RR15529.1
51. Cook, J. R. & Stefanski, L. A. Simulation-extrapolation estimation in parametric measurement error models. *J. Am. Statist. Assoc.* **89**, 1314-1328 (1994). https://doi.org/10.2307/2290994
52. Misumi, M., Furukawa, K., Cologne, J. B. & Cullings, H. M. Simulation-extrapolation for bias correction with exposure uncertainty in radiation risk analysis utilizing grouped data. *J. R. Stat. Soc. Ser. C-Appl. Stat.* **67**, 275-289 (2018). https://doi.org/10.1111/rssc.12225
27

**Supplement A.**

**Table A1. Assumed distribution of persons by radiation dose group, based in part on distribution of person years in the Japanese atomic bomb survivor Life Span Study[1]**

| Dose group | Central estimate of dose (Gy) | Scaled numbers of persons |
|---|---|---|
| 1 | 0.01 | 2591 |
| 2 | 0.1 | 334 |
| 3 | 0.5 | 438 |
| 4 | 1.5 | 102 |
| 5 | 2 | 6 |